\documentclass[12pt]{article}
\pdfoutput=1
\usepackage{amsmath,amssymb}
\usepackage{graphicx}
\begin{document}

\def\a{\alpha}
\def\b{\beta}
\def\c{\varepsilon}
\def\d{\delta}
\def\e{\epsilon}
\def\f{\phi}
\def\g{\gamma}
\def\h{\theta}
\def\k{\kappa}
\def\l{\lambda}
\def\m{\mu}
\def\n{\nu}
\def\p{\psi}
\def\q{\partial}
\def\r{\rho}
\def\s{\sigma}
\def\t{\tau}
\def\u{\upsilon}
\def\v{\varphi}
\def\w{\omega}
\def\x{\xi}
\def\y{\eta}
\def\z{\zeta}
\def\D{\Delta}
\def\G{\Gamma}
\def\H{\Theta}
\def\L{\Lambda}
\def\F{\Phi}
\def\P{\Psi}
\def\S{\Sigma}

\def\o{\over}
\def\beq{\begin{eqnarray}}
\def\eeq{\end{eqnarray}}
\newcommand{\gsim}{ \mathop{}_{\textstyle \sim}^{\textstyle >} }
\newcommand{\lsim}{ \mathop{}_{\textstyle \sim}^{\textstyle <} }
\newcommand{\vev}[1]{ \left\langle {#1} \right\rangle }
\newcommand{\bra}[1]{ \langle {#1} | }
\newcommand{\ket}[1]{ | {#1} \rangle }
\newcommand{\EV}{ {\rm eV} }
\newcommand{\KEV}{ {\rm keV} }
\newcommand{\MEV}{ {\rm MeV} }
\newcommand{\GEV}{ {\rm GeV} }
\newcommand{\TEV}{ {\rm TeV} }
\def\diag{\mathop{\rm diag}\nolimits}
\def\Spin{\mathop{\rm Spin}}
\def\SO{\mathop{\rm SO}}
\def\O{\mathop{\rm O}}
\def\SU{\mathop{\rm SU}}
\def\U{\mathop{\rm U}}
\def\Sp{\mathop{\rm Sp}}
\def\SL{\mathop{\rm SL}}
\def\tr{\mathop{\rm tr}}

\def\IJMP{Int.~J.~Mod.~Phys. }
\def\MPL{Mod.~Phys.~Lett. }
\def\NP{Nucl.~Phys. }
\def\PL{Phys.~Lett. }
\def\PR{Phys.~Rev. }
\def\PRL{Phys.~Rev.~Lett. }
\def\PTP{Prog.~Theor.~Phys. }
\def\ZP{Z.~Phys. }


\baselineskip 0.7cm

\begin{titlepage}

\begin{flushright}
IPMU11-0131
\end{flushright}

\vskip 1.35cm
\begin{center}
{\large \bf
Probing Extra Matter in Gauge Mediation
\\
Through the Lightest Higgs Boson Mass
}
\vskip 1.2cm
Jason L. Evans${}^{1}$, Masahiro Ibe${}^{1,2}$ and  Tsutomu T. Yanagida${}^{1}$
\vskip 0.4cm

${}^{1}${\it Institute for the Physics and Mathematics of the Universe (IPMU),\\ University of Tokyo, Chiba, 277-8583, Japan}\\
${}^{2}${\it Institute for Cosmic Ray Research, University of Tokyo, Chiba 277-8582, Japan }

\vskip 1.5cm

\abstract{
We discuss the implications of
the excesses in LHC Higgs boson searches on the
gauge mediated supersymmetric standard model, for the mass range $120-140$\,GeV.
We find that a relatively heavy lightest Higgs boson mass in this range
can be reconciled with light SUSY particles, $m_{\rm gluino}<2$\,TeV,
if there are additional fields which couple directly to the Higgs boson.
We also find that the mass
of this extra matter can be predicted rather precisely in gauge mediation for a given Higgs boson and gluino mass.
}
\end{center}
\end{titlepage}

\setcounter{page}{2}

\section{Introduction}
Models with gauge mediated supersymmetry  (SUSY) breaking\,%
\cite{Dine:1981za, Dine:1981gu, Dimopoulos:1982gm,Affleck:1984xz,Dine:1993yw,
Dine:1994vc,Dine:1995ag}
are one of the most attractive and phenomenologically viable realizations of the minimal supersymmetric standard model (MSSM).
One widely acknowledged prediction of gauge mediation is that
the lightest Higgs boson mass is near the edge of the presently allowed mass region. This small Higgs mass is due to a nearly vanishing $A$ terms. For example, the lightest Higgs boson mass is predicted to be smaller than $120$\,GeV
for $m_{\rm gluino} \lesssim 2$\,TeV.

The ATLAS and CMS collaborations at the LHC experiment, on the other hand, have
recently reported excesses in the Higgs boson searches in the mass range
$m_{h^0} = 120-140$\,GeV with a confidence level close to 3$\sigma$\,\cite{LHC}.
This appears to be troubling news for gauge mediation since a Higgs mass this size would require very heavy SUSY particles.
If the lightest Higgs boson is $\sim 140\,$GeV,
the gluino mass of gauge mediation would be far beyond the reach of the LHC experiments, i.e. $m_{\rm gluino}\gg 2\,$TeV.

In this paper, we point out that such seeming tension between the Higgs search
and the SUSY search in gauge mediation follows from the assumption that
there is no additional matter beyond the MSSM.
In contrast to such views, we claim that such a large Higgs boson mass suggests the existence
of additional fields which couple to the Higgs boson.
In fact in this scenario we will see that the lightest Higgs boson can be heavier than $140$\,GeV
even for $m_{\rm gluino}\lesssim 2$\,TeV.%
\footnote{
The enhancement of the lightest Higgs boson mass was discussed in
Refs.\,\cite{MO,Babu:2008ge,Martin:2009bg}.
}

The organization of the paper is as follows.
In section\,\ref{sec:constraints}, we discuss the current constraints on the mass and the coupling
constants of the additional matter fields.
In section\,\ref{sec:Higgs}, we discuss the effects of this additional matter on the lightest
Higgs boson mass.
The final section is devoted to discussions and conclusions.

\section{Models with extra matter}\label{sec:constraints}
In most gauge mediation models, the messengers are assumed to
be a fundamental and anti-fundamental of the minimal grand unified gauge
group, $SU(5)$.
In this way, gauge mediation is consistent with grand unified theories (GUT),
which seem very likely given that the gauge couplings unify in the MSSM.
It is important to note that there is room
for extra low energy matter multiplets beyond the above mentioned messenger.
That is, keeping perturbative unification at the GUT scale,
we can still add
\begin{itemize}
\item up to one pair of $\mathbf {\bar{5} + 10}$
\item up to three pairs of $\mathbf {5 + \bar{5}   }$
\item up to one pair of $\mathbf {10 + \bar{10} }$
\end{itemize}
along with an arbitrary number of gauge singlets.%
\footnote{
In this paper, we assume that the extra matter
forms a complete multiplet.
}

In the above lists, the first set of fields corresponds to a fourth-generation of matter.
The CMS collaboration, however, has
ruled out a fourth generation for a Higgs boson mass in the range $120-600$\,GeV
at a 95\%\,C.L. due to the Higgs production cross section being too large\,\cite{SM4}.
Since we are interested in the Higgs boson mass in the range of $120-140$\,GeV,
we do not pursue this possibility, but rather concentrate on the other two possibilities with vector-like multiplets%
\footnote{
We may allow a pair of $\mathbf{\bar{5}+10}$ and $\mathbf{{5}+\bar{10}}$
keeping the perturbative unification when the messenger scale is very high.
This set of extra matter corresponds to a vector-like fourth generation of matter.
}
(see discussion on the Higgs production cross section
in the case of vector-like extra matter in the appendix\,\ref{sec:HIGLU}).
We assume that the masses of these vector like fields are similar to the Higgsino masses,
i.e. the so-called $\mu$-term, otherwise their masses should naturally be at the GUT scale.

The addition of a $\mathbf{ 10 + \bar{10}}$ is particularly interesting,
since the doublet Higgs $H_u$ may couple to this extra matter via $W= H_u \mathbf{10\, 10}$.
From this additional Higgs interaction, the lightest Higgs boson mass can
be increased due to radiative corrections\,\cite{MO}.
The $\mathbf{5 +\bar{5}}$ mentioned above can also couple to $H_u$
if we also introduce a singlet ${\mathbf 1}$, $W = H_u\, {\mathbf{\bar 5}}\, \mathbf 1 $.
In the following, however, we mainly concentrate on the model with an extra
$\mathbf {10 + \bar{10}}$. This matter content is more advantageous since 
the gauge mediated contribution to the soft masses of the $\mathbf {10 + \bar{10}}$ are much larger.

\subsection{Constraints on extra matter}
Here we concentrate on the constraints on the additional matter,
\begin{eqnarray}
{\mathbf {10}} = (Q_E, \bar{U}_E, \bar{E}_E) \ ,\quad
{\mathbf { \bar{10}}} = (\bar{Q}_E, {U}_E, E_E) \ ,
\end{eqnarray}
where their Standard Model charge assignments should be clear.
In terms of these fields, the Yukawa interaction $W= H_u \mathbf{10\, 10}$
and invariant mass term $W= M_T \mathbf{10\,\bar{10}}$ can be rewritten as,
\begin{eqnarray}
\label{eq:super1}
   W =M_T \bar{Q}_EQ_E + M_T \bar{U}_EU_E  + M_T \bar{E}_EE_E
   +  \lambda_u H_u Q_E \bar{U}_E\ ,
\end{eqnarray}
where $\lambda_u$ denotes a coupling constant.
As a result of the Higgs coupling, the masses of the additonal up-type quarks
are split,
\begin{eqnarray}\label{eq:uptype}
 M_{\rm up-type} \simeq M_T \pm \frac{\lambda_u\sin\beta v}{2}\ ,
\end{eqnarray}
with the Higgs vacuum expectation value $v \simeq 174$\,GeV.
The masses of the other extra matter multiplets remain degenerated and are given by $M_T$.

As we will show in the next section, the lightest Higgs boson
mass is considerably enhanced for $\lambda_u \simeq 1$.
From this point on, we will assume that $\tan\beta = \vev{H_u}/\vev{H_d} = O(10)$,
since we are interested in models with a relatively heavy lightest Higgs boson.%
\footnote{In our discussion, we could also include another Yukawa interaction
$W= \lambda_d H_d \mathbf{\bar{10}\, \bar{10}}$.
This interaction, however, does not change our discussion
even for $\lambda_d \simeq \lambda_u$ as long as $\tan\beta = O(10)$. }

In the superpotential in Eq.\,(\ref{eq:super1}),
we have implicitly assumed that the extra matter does not
mix with the matter multiplets of the MSSM. These couplings can be forbidden
by assuming some type of parity for which the extra matter is odd.
If the mixing between the MSSM and the extra matter are completely
forbidden, however,
some of the extra charged particles can be stable, which
conflicts with cosmological constraints.
To avoid such difficulties, we assume
that the additional matter has a small amount of mixings with the MSSM matter multiplets
($\mathbf{\bar{5}}_{i=1-3}, \mathbf{10}_{i=1-3}$), such as
\begin{eqnarray}
 W = \epsilon_i H_u \mathbf{ 10_i 10 } + \epsilon_i H_d\mathbf{ \bar{5}_i  10}.
\end{eqnarray}
Even with this small additional mixing the extra matter in $\mathbf{ 10}$ can
decay into the MSSM fields. The  $\mathbf{\bar{10}}$ field can
also decay into the MSSM fields through the mass terms in Eq.\,(\ref{eq:super1}).%
\footnote{
The amplitudes of the FCNC process such as $K^0-\bar{K}^0$ mixing
caused by these terms are suppressed by $\epsilon^4$.
Thus, these terms do not cause FCNC problems
for $\epsilon \lesssim 10^{-3}$.
}
Particulary, the lightest additional colored particle, which is an up-type quark, can
decay into a $W$-boson and a down-type quark.

The most stringent collider constraint on the mass of these heavy up-type quarks is
\begin{eqnarray}
M_{\rm up-type} > 358\,{\rm GeV}\ .
\end{eqnarray}
The above constraint is for up-type quarks decaying to a $W$-boson and a down type quark and was obtained by the CDF collaboration at the Tevatron experiments
using $5.3$\,fb$^{-1}$ of data in the channel lepton+jets\,\cite{CDF}.%
\footnote{
Using $37\,{\rm pb}^{-1}$ data set, the ATLAS collaboration
placed a constraint on a heavy quark decaying to
a $W$-boson and quarks giving $M_T>270$\,GeV\,\cite{ATLAS}.
The CMS collaboration has also reported more stringent
constraints on the up-type fermion masses from examining decays to a $W$-boson
and bottom quark using $573-821\,{\rm pb}^{-1}$, which gives $M_T > 450$\,GeV.
}

Using the mass constraints found in Eq.\,(\ref{eq:uptype}),
we find that collider constraints place a rough lower bound on $M_T$,
\begin{eqnarray}
M_T \gtrsim 445\,{\rm GeV}\ .
\end{eqnarray}
As for the decays of extra scalars, they place much weaker constraints on $M_T$,
since their production cross section is suppressed compared
with the extra fermions.%
\footnote{The collider constraint on $M_{\rm up-type}$
can be weaken by assuming the mixing between the extra matter
and the MSSM matter is very small, so that the extra colored matter would have a rather long lifetime.
In this case, a smaller $M_T$ is allowed.}

The new fields which couple to the Higgs boson also give
important contributions to the electroweak precision observables, $S$, $T$ and $U$\,\cite{Peskin:1990zt}.
To see such contributions, we compute the $S$, $T$ and $U$ parameters
at the one-loop level and find
\begin{eqnarray}\label{eq:STU}
S &\simeq& \frac{2}{5\pi}\frac{(\lambda_u \sin\beta v)^2}{M_T^2}\ , \cr
T &\simeq& \frac{13}{160\pi s_W^2 c_W^2}\frac{(\lambda_u \sin\beta v)^4}{M_T^2m_Z^2}\ , \cr
U &\simeq& \frac{13}{140\pi }\frac{(\lambda_u \sin\beta v)^4}{M_T^4}\ ,
\end{eqnarray}
where $s_W^2 \simeq 0.231$ and $c_W^2 = 1-s_W^2$ denote the weak mixing angle.
(See also the appendix\,\ref{sec:STU} and Ref.\,\cite{Martin:2009bg} for detailed analysis.)
The above results neglect contributions from the additional scalar fields,
since they are expected to be heavier than their fermion counterpart.

\begin{figure}[t]
\begin{center}
  \includegraphics[width=.5\linewidth]{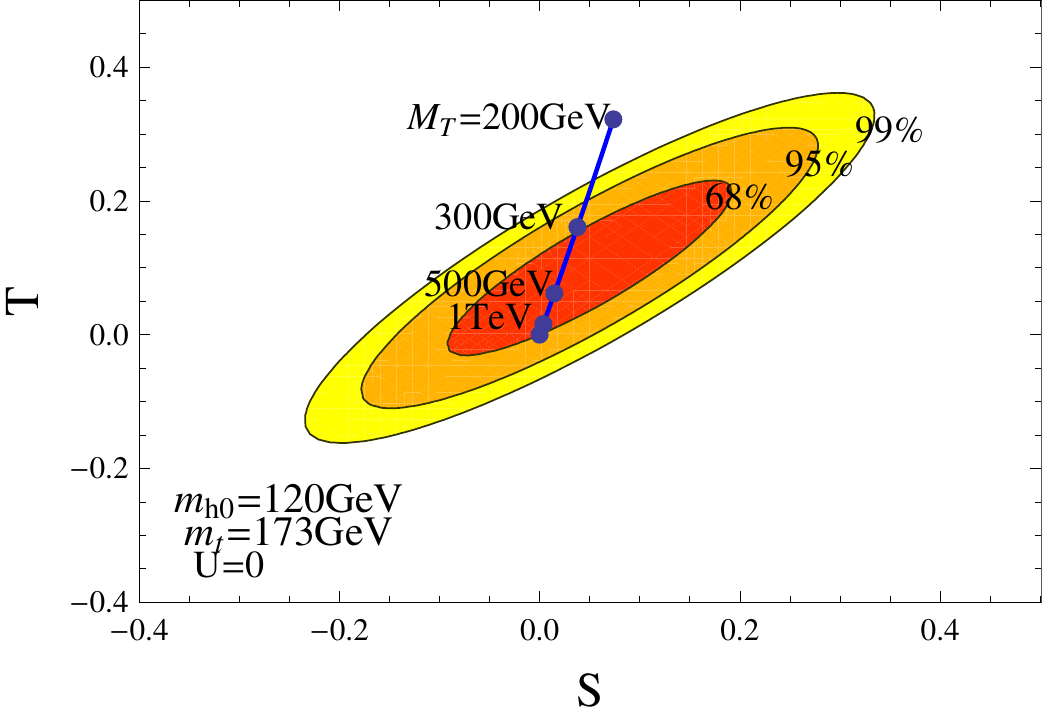}
\caption{\sl \small
The contributions to the electroweak precision
observables $S$ and $T$ for $\lambda_u = 1$ from the additional $\mathbf{ 10 + \bar{10}}$.
In this figure, we have neglected the contributions from
the extra scalar fields assuming that they are heavier than
about a TeV.
The ellipses of $68$\%, $95$\% and $99$\% C.L.
in Ref.\,\cite{Baak:2011ze} are also shown.
}
\label{fig:ST}
\end{center}
\end{figure}

In Fig.\,\ref{fig:ST}, we show the shift in the $S$ and $T$-parameters due to the additional matter
for $\lambda_u = 1$.%
\footnote{
The coupling $\lambda_u =1$ can be perturbative up to the GUT scale
due to the quasi-fixed point behavior\,\cite{Martin:2009bg}.
}
In this figure, we also show the electroweak precision constraints on new physics at
the $68$\%, $95$\% and $99$\% C.L.
for $m_{h^0}=120$\,GeV, $m_{\rm top} = 173$\,GeV and $U=0$
given in Ref.\,\cite{Baak:2011ze}.
The figure shows that contributions to the $S$ and $T$ parameters from the extra matter satisfy the experimental constraints for
$M_T\gtrsim 270$\,GeV at 95\% C.L..
Moreover, the contribution from extra matter makes the fit to the $Z$-pole data better
than the Standard Model as long as
\begin{eqnarray}
300\,{\rm GeV} \lesssim M_T \lesssim 1\,{\rm TeV}\ .
\end{eqnarray}
As we will see in the next section, it is remarkable that this mass range is
the preferred region for a lightest Higgs boson mass around $m_{h^0} \simeq 130-140$\,GeV
for $m_{\rm gluino} < 2$\,TeV.

\section{The lightest Higgs boson mass}\label{sec:Higgs}
The scalar potential of the Higgs boson receives additional radiative corrections
from the new interactions in Eq.\,(\ref{eq:super1}).
Most importantly, the effective quartic term of the Higgs boson,
\begin{eqnarray}
V= \frac{\lambda_{\rm eff}}{2} |H_u|^4\ ,
\end{eqnarray}
receives a positive contribution from its interactions with this extra matter if the extra scalar fields are heavier than their fermionic counterpart.

Calculating the one-loop four-point diagrams, we obtain the effective quartic coupling,
\begin{eqnarray}
\label{eq:effectiveQ}
{\mit \Delta} \lambda_{\rm eff} &\simeq&
 \frac{\lambda_u^4}{16\pi^2 (x_Q+1)^2}(6 (x_Q + 1)^2 \log(x_Q + 1) - x_Q (5 x_Q + 4))\ ,\cr
 & \simeq & \frac{\lambda_u^2}{16\pi^2} (6 \log x_Q - 5)\ , \quad (x_Q \gg 1)\ .
\end{eqnarray}
Here, $x_{Q}$ is defined by,
\begin{eqnarray}
   x_{Q} = \frac{m_{Q}^2}{M_T^2}\ ,
\end{eqnarray}
where we have assumed degenerate gauge mediated SUSY breaking squared masses for the up-type squarks, $m_{Q}^2 \simeq m_{U}^2$. We have also neglected $A$-term contributions to the effective quartic term,
since they are generically small in models with minimal gauge mediation.

With these additional contributions to the effective quartic coupling, the lightest Higgs boson mass is,
\begin{eqnarray}
m_{h^0}^2 \simeq m_Z^2 \cos^22\beta
 + \frac{3}{4\pi^2}
y_t^2 m_t^2 \sin^2\beta
\log\frac{m_{\tilde t}^2}{m_t^2}
+ \frac{4 {\mit \Delta}\lambda_{\rm eff}}{(g^2 + g'^2)} m_Z^2 \sin^2\beta
+ \cdots
\ ,
\end{eqnarray}
where the second term comes from
the one-loop top-stop contributions\,\cite{Higgs}, the third term is the new contribution calculated above,
and the ellipses denote the higher-loop contributions.

\begin{figure}[t]
\begin{center}
\begin{minipage}{.49\linewidth}
  \includegraphics[width=.9\linewidth]{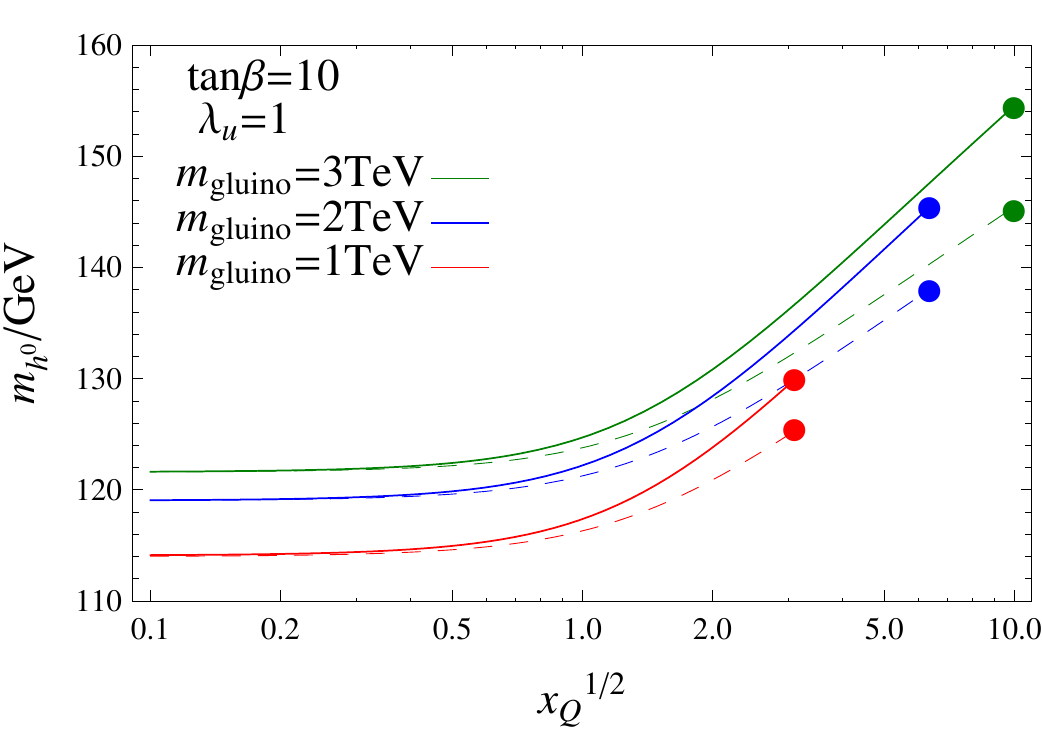}
  \end{minipage}
  \begin{minipage}{.49\linewidth}
  \includegraphics[width=.9\linewidth]{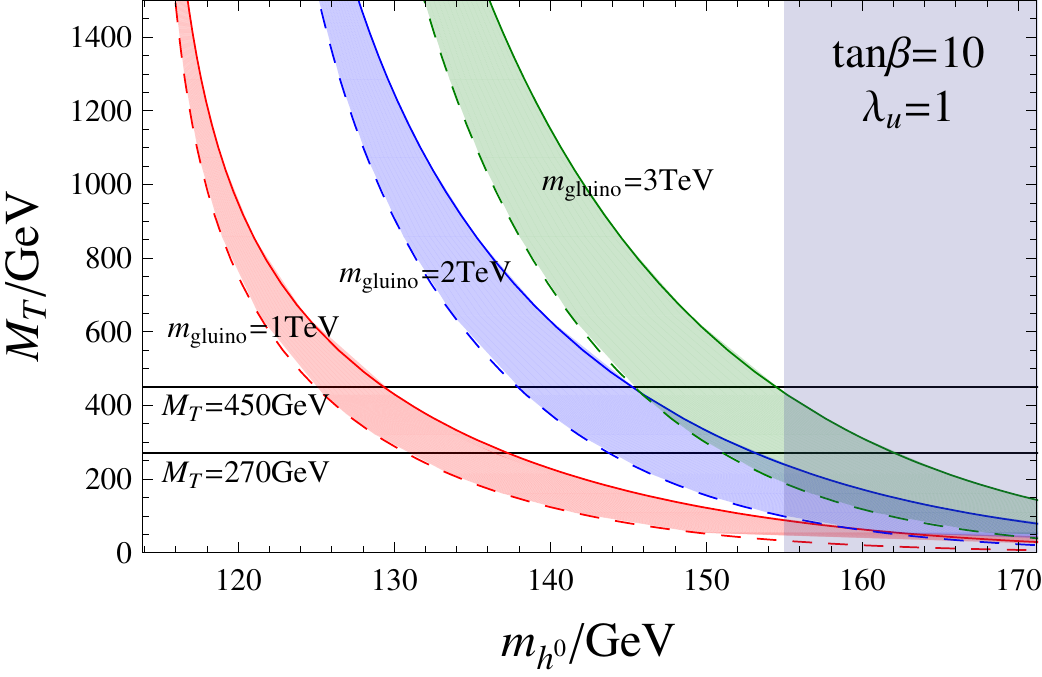}
  \end{minipage}
\caption{\sl \small
Left)
The lightest Higgs boson mass for a given $x_Q$.
In this figure we have taken $\tan\beta = 10$ and $\lambda_u = 1$.
The upper bounds on $x_Q$ of each line correspond to
$M_T = 450$\,GeV.
The region $x_Q \ll 1$ corresponds to the MSSM limit.
The dashed lines denote the Higgs boson mass
with the effective quartic term suppressed by $30$\% to simulate the two-loop contribution.
Right)
The required value of $M_T$ for a given lightest Higgs boson mass.
The upper and the lower boundaries of each band
correspond to the required $M_T$
based on the one-loop quartic coupling in Eq.\,(\ref{eq:effectiveQ})
and the one suppressed by 30\% which is again to simulate the two-loop contributions.
The dark shaded region corresponds to
lightest Higgs boson masses that are excluded by the LHC experiments
at 95\%C.L.\cite{LHC}.
}
\label{fig:Higgs}
\end{center}
\end{figure}

In Fig.\,\ref{fig:Higgs}, we show the Higgs boson mass
for a given value of the gluino pole mass and $x_Q$.
In the figure, we have taken $\lambda_u = 1$.
To obtain the mGMSB contribution to the lightest Higgs boson mass,
for a given gluino mass, we used
{\it SoftSusy}\,\cite{Allanach:2001kg}.
The upper bounds on $x_Q$, for each line in Fig.\,\ref{fig:Higgs}, corresponds to choosing $M_T\simeq 450$\,GeV, which is  rough the experimental lower bound.%
\footnote{
In our analysis, we have neglected the effects of $\lambda_u$ on the running
of the soft SUSY breaking mass $m_Q$ and taken $m_Q = (1.3, 2.8, 4.5 )$\,TeV
for $m_{\rm gluino}  = (1,2,3)$\,TeV, respectively.
}

The figure shows that the Higgs boson can be
much heavier than $140$\,GeV even for $m_{\rm gluino } < 2$\,TeV.
Notice that we have not included two-loop contributions
from the extra matter fields in our analysis.
In the case of the stop-loop contributions, for example,
the two-loop contributions reduce the radiative
corrections to the effective Higgs quartic term
by about $30$\%\,\cite{Higgs}.
Although the two-loop corrections to the Higgs boson mass
from the extra matter goes beyond the scope of our paper,
we show the lightest Higgs boson mass with
the additional contribution to one-loop effective quartic term
suppressed by 30\% for comparison.
With this suppression, we still find that the these extra matter fields
can push the mass of the lightest Higgs boson above $140$\,GeV
for $m_{\rm gluino} <2$\,TeV.

In the right panel of Fig.\,\ref{fig:Higgs}, we also plot the required
invariant mass of the extra matter as a function of
the lightest Higgs boson mass.
The figure shows that the mass of
the extra matter can be predicted rather precisely
for a given Higgs boson mass.
For example, the lightest Higgs boson mass of $130-140$\,GeV
requires the existence of additional matter with mass less than about a TeV
for $m_{\rm gluino} \lesssim 2$\,TeV.
It is also interesting to note that some of the parameter space
has already been excluded by the LHC experiments
because the lightest Higgs mass is too large i.e. $155\,{\rm GeV}<m_{h^0}<190$\,GeV.

\section{Discussion and conclusion}
In this paper, we discussed
the implications of the recently reported excesses in the Higgs boson searches at the LHC experiments
in the mass range $m_{h^0} = 120-140$\,GeV for models with gauge mediation.
As we have seen, such a relatively heavy lightest Higgs boson
can be reconciled with light SUSY particles, i.e. $m_{\rm gluino}<2$\,TeV,
if there is extra matter that couples to the Higgs doublet.
Furthermore, due to the robust prediction of the lightest Higgs boson
mass in gauge mediation models, we find that
the mass of the extra matter can be predicted rather precisely
for a given a Higgs boson and gluino mass.%
\footnote{To make more precise prediction,
it is important to include two-loop extra matter contributions to the
lightest Higgs boson mass.}
It is also remarkable that the invariant mass of the extra matter
preferred by a Higgs boson mass in the range $m_{h^0} \simeq 130-140$\,GeV
gives a better fit to the electroweak precision measurements as seen in Fig.\,\ref{fig:ST}.

Finally, let us comment on the effects of the $A$-term.
In this paper, we have neglected the $A$-term contributions
to the lightest Higgs boson mass,
since the $A$-terms are predicted to be quite small
in minimal gauge mediation models.
As recently discussed in Ref.\,\cite{Asano:2011zt}, however,
the lightest Higgs boson can be significantly increased
even if the extra matter is relatively heavy if large $A$-terms are generated.
In fact, it is possible to generate large $A$-terms
in models with gauge mediation as is discussed in Ref.\,\cite{Evans:2011be}.
In cases with large $A$-terms, the predicted mass for the extra matter
in this paper will change.%
\footnote{
The analysis of models with extra matter with large $A$-terms
will be given elsewhere.
}

\section*{Note added}
Within a few days of completing this manuscript,
Ref.\,\cite{Endo:2011mc}
was posted on arXiv, which overlaps with some of our discussion
on the lightest Higgs boson mass in gauge mediation.
\section*{Acknowledgements}
We would like to thank K.Harigaya for useful discussion.
This work was supported by the World Premier International Research Center Initiative
(WPI Initiative), MEXT, Japan.
The work of T.T.Y. was supported by JSPS Grand-in-Aid for Scientific Research (A)
(22244021).

\appendix
\section{Higgs Effective couplings to gluons}\label{sec:HIGLU}
In the limit where the additional fermions are much heavier than
the other energy scales in the problem,
the Higgs effective couplings to gluons can be derived from the
effective operator,
\begin{eqnarray}
{\cal L}_{\rm eff} = \frac{\alpha_s}{8\pi}\frac{1}{3}
\left.\frac{\partial}{\partial h}\log \det{\cal M}(h)\right|_{h=0}\times h\,
 G^a_{\mu\nu}G^{a\mu\nu}\ .
\end{eqnarray}
The effective mass matrix ${\cal M}(h)$ is given by,
\begin{eqnarray}
{\cal M} =
\left(
\begin{array}{cc}
M_T    & \lambda_u \sin\beta \, v  \\
\lambda_d \cos\beta\, v    &   M_T
\end{array}
\right)
\end{eqnarray}
by replacing $v \to v + h/\sqrt{2}$.%
\footnote{We have taken the so called decoupling limit
for the heavier Higgs bosons.}
Here, we have included the effects of the Yukawa coupling,
\begin{eqnarray}
W = \lambda_d H_d \bar{Q}_E U_E\ ,
\end{eqnarray}
which was neglected in the main text.
In the limit of $M_T \gg v$, we obtain,
\begin{eqnarray}
{\cal L}_{\rm eff} = \frac{\alpha_s}{8\pi}\frac{1}{3}
\frac{\lambda_u\lambda_d \sin2\beta v^2}{M_T^2}
\frac{ h}{\sqrt{2} v}\,
 G^a_{\mu\nu}G^{a\mu\nu}\ .
\end{eqnarray}

Therefore, compared with the top loop contribution to the effective gluon coupling,
 \begin{eqnarray}
{\cal L}_{\rm eff}^{\rm top} = \frac{\alpha_s}{4\pi}\frac{1}{3}
\frac{ h}{\sqrt{2} v}\,
 G^a_{\mu\nu}G^{a\mu\nu}\ ,
\end{eqnarray}
we find that the contribution to the Higgs effective couplings to gluons
is negligible for,
\begin{eqnarray}
  M_T^2 \gg \lambda_u\lambda_d \sin2\beta v^2\ .
\end{eqnarray}

In this discussion, we neglected the
the extra scalar particles.
The scalar contributions are, however, further suppressed
by a large SUSY breaking masses,
and hence, they can be safely neglected.%
\footnote{The scalar contributions to the Higgs effective coupling to gluons
can be obtained by replacing $\det{\cal M}$ to the mass matrix of scalar component, i.e.
$\det{\cal M}_B^2$.}

\section{Precision Electroweak Parameters}\label{sec:STU}
In this appendix, we discuss contributions of the extra matter
to the electroweak precision observables, $S$, $T$ and $U$,
which are defined by\,\cite{Peskin:1990zt},
\begin{eqnarray}
\alpha S &=& 4 s_W
c^2_W
\left[
\Pi_{ZZ}'(0)
-\frac{c^2_W-s^2_W}{s_Wc_W}
\Pi_{Z\gamma}'(0)
-
\Pi_{\gamma\gamma}'(0)
\right]\ ,
\cr
\alpha T &=& \frac{\Pi_{WW}(0)}{M_W^2} -\frac{\Pi_{ZZ}(0)}{M_Z^2}\  ,
\cr
\alpha U &=& 4s_W^2
\left[
\Pi_{WW}'(0)
-c_W^2 \Pi_{ZZ}'(0)
-2 s_W c_W \Pi_{Z\gamma}'(0)
-s_W^2 \Pi_{\gamma\gamma}'(0)
\right]\ .
\end{eqnarray}
The dominant contributions to the electroweak precision observables, from our additional vector like matter,
comes from the fermionic components of the up-type quarks
whose mass matrix is given by,
\begin{eqnarray}
{\cal M} =
\left(
\begin{array}{cc}
M_T    & \lambda_u \sin\beta \, v  \\
0    &   M_T
\end{array}
\right)\ ,
\end{eqnarray}
where we have again omitted $W = H_d \mathbf{\bar{10}\,\bar{10}}$.

By computing the one-loop contributions to the vacuum polarization,
we obtain the follow corrections to $S$, $T$  and $U$,
\begin{eqnarray}
S &=& \frac{1}{6\pi} \frac{1}{(M_H^2-M_L^2)^3(M_H+M_L)^2}
\bigg[
3 M_H^8-10 M_H^7 M_L-24 M_H^6 M_L^2+54 M_H^5 M_L^3-54 M_H^3 M_L^5
+24 M_H^2 M_L^6
\cr
&&
+10 M_H M_L^7-3 M_L^8 \left.
+6 M_H M_L (M_H^6 - 3 M_H^4 M_L^2 + 6 M_H^3 M_L^3 - 3 M_H^2 M_L^4 + M_L^6)
\log\left(
\frac{
  M_H^2}{M_L^2}
  \right)\right] \ ,
\end{eqnarray}
\begin{eqnarray}
T &=& \frac{3}{16\pi s_W^2 c_W^2} \frac{1}{m_Z^2(M_H^2-M_L^2)(M_H+M_L)^2}
\bigg[
M_H^6-6 M_H^5 M_L-3 M_H^4 M_L^2+3 M_H^2 M_L^4+6 M_H M_L^5-M_L^6
\cr
&&\left.
+ M_H M_L (M_H^4 + 10 M_H^2 M_L^2 + M_L^4)
\log\left(
\frac{
  M_H^2}{M_L^2}
  \right)\right] \ ,
\end{eqnarray}
\begin{eqnarray}
U &=& \frac{1}{6\pi } \frac{1}{m_Z^2(M_H^2-M_L^2)(M_H+M_L)^2}
\bigg[
-7 M_H^8-10 M_H^7 M_L-16 M_H^6 M_L^2-90 M_H^5 M_L^3
+90 M_H^3 M_L^5\qquad
\cr
&&\qquad \qquad \qquad\qquad\qquad\qquad\qquad\qquad\qquad\qquad\qquad
\qquad\quad
+16 M_H^2 M_L^6+10 M_H M_L^7+7 M_L^8
\cr
&&\left.
+6 (M_H^8 + 4 M_H^6 M_L^2 + 20 M_H^5 M_L^3 + 10 M_H^4 M_L^4 + 20 M_H^3 M_L^5 +
   4 M_H^2 M_L^6 + M_L^8) \log\left(
\frac{
  M_H^2}{M_L^2}
  \right)\right] \ .
\end{eqnarray}
Here, the masses $M_{H,L}$ are given by,
\begin{eqnarray}
M_{H,L}^2 = M_{T}^2 + \frac{(\lambda_u\sin\beta\, v)^2}{2}
\pm \frac{\lambda_u\sin\beta\, v}{2}
\sqrt{4 M_T^2 + (\lambda_u \sin\beta \,v)^2}\ ,
\end{eqnarray}
which satisfy,
\begin{eqnarray}
 M_T^2 = M_H M_L \ .
\end{eqnarray}
We have also taken the phase convention $M_T > 0$, $\lambda_u \sin\beta \, v>0$.
By taking the limit where $(\lambda_u \sin\beta v)/M_T$ is small,
we obtain the $S$, $T$ and $U$ parameters in Eqs.\,(\ref{eq:STU}).

\end{document}